\documentclass[noshowpacs, nofootinbib, twocolumn]{revtex4-1}

\usepackage{amsmath, amssymb, graphicx}
\usepackage[hyperindex=true,
          pdfstartview=FitH,
          bookmarksnumbered=true,
          bookmarksopen=true,
          citecolor=blue,
          linkcolor=blue,
          colorlinks=true, 
          pdfborder=001,   
          unicode]{hyperref}

\begin{document}

\title{Hidden symmetries for thermodynamics and emergence of relativity}
\author{Liu Zhao}
\affiliation{School of Physics, Nankai University, Tianjin 300071, P R China
\\{email: {\it lzhao@nankai.edu.cn}}}
\date{\today}

\begin{abstract}
Erik Verlinde recently proposed an idea about the thermodynamic origin of
gravity. Though this is a beautiful idea which may resolve many long
standing problems in the theories of gravity, it also raises many other
problems. In this article I will comment on some of the problems of 
Verlinde's proposal with special emphasis on the thermodynamical origin 
of the principle of relativity. It is found that there is a large group of hidden 
symmetries of thermodynamics which contains the Poincare group of the 
spacetime for which space is emergent. This explains the thermodynamic 
origin of the principle of relativity.
\end{abstract}
\maketitle

\section{Introduction}

Of all physical theories about Nature, only two branches are 
independent of concrete details of the systems being considered, i.e. 
thermodynamics and relativity (in either the special or general sense). These 
two theories are about the universal principles every 
physical system must obey and are hence referred to as principle 
theories. All other theories are about concrete systems. It is puzzling why in 
the first place Nature prefers to abide by {\em two principle theories} rather 
than simply one (puzzle of two), because two is more dangerous in the 
sense that potential contradictions between them might occur. However, 
Nature is quite smart in avoiding such contradictions. Even more, in many 
cases the two principle theories seem to be mysteriously connected 
to each other. This connection is especially transparent when one talks 
about relativity in the general sense: on the one hand, for every black hole 
solution of the general theory of relativity one can define thermodynamic 
quantities like entropy and temperature and check that the classical laws of 
thermodynamics are obeyed; on the other hand, it has repeatedly been 
discussed that the Einstein equation \cite{Jacobson:1995p4065} 
\cite{Padmanabhan:2009p5086} \cite{Padmanabhan:2009p5122} and the 
action \cite{Caravelli:2010p5148} of the general theory of relativity can be 
derived from the first law of thermodynamics, and recently Verlinde 
\cite{Verlinde:2010p5107} went 
further by proposing that gravity is an entropic force and thus calling for an 
end of gravity as a fundamental force. Subsequent surge of works appeared 
almost instantly, including \cite{Smolin:2010p5123}, which applied 
Verlinde's idea in loop quantum gravity, 
\cite{Padmanabhan:2010p5128} and \cite{Cai:2010p5127}, both derived 
the Friedman equation using Verlinde's idea,  
\cite{Wang:2010p5132}, which applied Verlinde's idea in establishing 
UV/IR relation and obtaining holographic dark energy,  
\cite{Wang:2010p5136}, which tries to interpret electrostatic force also as an 
entropic force,  \cite{Zhang:2010p5140},  \cite{Ling:2010p5141} 
and \cite{Wei:2010p5145}, which applied Verlinde's ideas in 
the settings of modified gravity, brane cosmology and Horava gravity 
respectively,  and \cite{Wang:2010p5144}, which proposed two novel 
kind of cosmic perturbations corresponding to Verlinde's emergent gravity. 
The idea of emergent gravity also attracted some quantum 
information people, see \cite{Lee:2010p5153} for an alternative proposal.

Verlinde's proposal is certainly very beautiful. If proven true, a significant 
portion of modern theories of fundamental physics should be reformulated. 
It not only has the potential of addressing the long standing problem of 
quantum gravity by identifying gravity as a purely macroscopic effect, but 
also implies that space itself must be emergent as a macroscopic effect. 
Moreover, it has the potential to resolve the puzzle of two mentioned 
above. However, Verlinde's idea also raises many new problems because 
most concepts taking space as a fundamental existence must be 
reexamined. Verlinde himself has 
noticed this point and explained in particular the concepts of inertia and 
Newton's law of classical mechanics from the new point of view. 

In this article I will comment that there are more problems which should 
be checked against before one can take Verlinde's proposal more seriously. 
The origin of the principle of relativity is 
among these problems. It is found that there is a large group of hidden 
symmetries for thermodynamics which has not been discussed before. By 
incorporating this hidden symmetry group together with Verlinde's idea and 
the second law of thermodynamics, it is shown that the principle of 
relativity
arises naturally. Meanwhile, a clearer understanding about the role of 
holographic principle in Verlinde's proposal is given. It turns out that 
holography is a requirement of democracy between different scaling 
dimensions and the fact that space is three dimensional. 

\section{Verlinde's proposal: problems to be answered}

Verlinde's proposal contains the following key ingredients: 
\begin{itemize}
\item space is emergent,  the part of space which has not yet emerged
is enclosed by a holographic screen and the entropy is proportional to the 
area of the screen;

\item gravity is an entropic force just like any other generalized forces 
entering in the first law of thermodynamics. More concretely, gravity 
is caused by the change of entropy behind the holographic screen due to 
the emergence of space;

\item the temperature is either related to the acceleration of the observer via 
Unruh's law or related to the total energy of the system via equipartition 
law (the equipartition of energy on black hole horizons is discussed 
earlier by Padmanabhan in \cite{Padmanabhan:2003p5146}, see also 
\cite{Padmanabhan:2009p5086}), 
and the total energy of the system is also equal to the mass behind the 
holographic screen times $c^2$, i.e. $E=Mc^2$. Using Unruh's law for 
temperature leads to Newton's law of classical mechanics, and using 
equipartition law and $E=Mc^2$ gives rise to Newton's law of gravitational 
force.
\end{itemize}
A lot more have been discussed in Verlinde's paper 
\cite{Verlinde:2010p5107}, however besides 
the above new ingredients the rest arguments and sketchy derivations seem to 
have been discussed in or implied by earlier works since Jacobson's 
\cite{Jacobson:1995p4065}, in 
particular, the formal derivation of Einstein equation is now new.

Now let us take a more careful look at the new ideas listed above.  

First of all, given that space is emergent, one has to explain why is it look 
like what we perceive it. In particular, why is space three dimensional, or 
even why the dimension of space is an integer? This is a problem one has to 
answer in an emergent theory of space but these were not paid for a single 
word in Verlinde's paper;

Second, it is known since the 1930's \cite{Tolman:1934} that classical 
thermodynamics is 
incompatible with the special theory of relativity if spacetime is considered as 
a fundamental existence. However, in an emergent theory of space one has to 
see the principle of relativity also as an emergent consequence. It remains to 
check whether the emergence of the principle of relativity is possible or not
following Verlinde's proposal, and I think this is by now one of the most 
severe obstacle to be overcome before this new proposal can be more widely 
accepted in the physics community;

Third, in deriving the Newton's law of gravity, the equipartition law together 
with the relation $E=Mc^2$ were used as input. But the origin of the latter 
relation is only known from special relativity. It gives an impression that 
special relativity is provided as another fundamental assumption, but even 
though it is still strange that in order to obtain the nonrelativistic law for 
gravity one has to use the relativistic relation for the total energy. Also the 
Unruh temperature relation is a consequence of relativistic motion, while it is 
used in deriving the nonrelativistic law of classical mechanics. Moreover, 
the use of equipartition law also feels strange, because equipartition law 
depends on the microscopic details of the system while Newton's law doesn't. 

There are other issues which need more interpretations but I prefer not to  
concentrate on them and just emphasis on the problem of emergence of the 
principle of relativity. If the principle of relativity is put in by hand, it will 
seriously hurt the picture of the emergence of space, and the puzzle of two 
will not be resolved. On the contrary, if the 
principle of relativity is emergent, then not only the puzzle of two is 
resolved, it will also call for an improvement for the derivation of the 
Newton's law of gravity, or simply view Newton's law as the nonrelativistic 
limit of general relativity as usual while considering general relativity as an 
emergent theory about gravity.  In either way the puzzle of deriving 
nonrelativistic laws using relativistic formulas without taking a 
nonrelativistic limit should be avoided. 

In the next section I will propose a possible solution to the emergence of 
the principle of relativity by unraveling a large group of hidden symmetries 
of classical thermodynamics. Assuming that space is emergent and plays as extensive variables in the first law of thermodynamics, it will be seen that the Poincare group is naturally contained in the hidden symmetry group of thermodynamics, which gives an explanation of the principle of relativity as an emergent concept from thermodynamic geometry.

\section{Geometry and hidden symmetries of classical 
thermodynamics} 

There are quite a few different prescriptions of thermodynamics as a 
geometric system. Gibbs \cite{J.Gibbs} described the thermodynamic 
phase space as a contact manifold. Weinhold \cite{F. Weinhold} and 
Ruppeiner \cite{G. Ruppeiner} respectively described the geometry of the 
thermodynamic configuration space (i.e. the space of equilibrium states) as 
Riemannian geometries (with different metric choices). And recently H. 
Quevedo \cite{Quevedo:2006p1979} \cite{Quevedo:2007p1262} described 
the geometry of the thermodynamic phase space as a 
contact Riemannian geometry invariant under the group of Legendre 
transformations.

For my purpose it is tempting to describe the geometry of the 
thermodynamic phase space as a contact, Riemannian geometry whose pull 
back to the thermodynamic configuration space has a metric with 
Lorentzian signature. 

To begin with, let me describe the contact geometry of the thermodynamic 
phase space $\mathcal{T}$, following the spirits of Gibbs \cite{J.Gibbs}. 
The space $\mathcal{T}$ is a $2n+1$ dimensional contact
manifold with coordinates $(\Phi, E^a, I_a)$  ($a=1, 2, ..., n$), where 
$\Phi$ is a thermodynamic potential,  $E^a$ and $I_a$ are respectively 
extensive and intensive variables of the system, and a contact one form 
$\Theta = d\Phi - \sum_a I_a dE^a$ obeying $(d\Theta)^{\wedge n} 
\wedge \Theta \neq 0$ is needed in order to identify the contact structure 
of the manifold $\mathcal{T}$. The space 
of classical equilibrium thermodynamical states $\mathcal{E} = 
\mathrm{span \ of \ }\{E^a\}$ 
is embedded in $\mathcal{T}$ as a subspace by a smooth mapping
\[
\varphi: \mathcal{E} \rightarrow \mathcal{T}, \qquad
\varphi (E^a) = (\Phi(E_a), E^a, I_a(E^a)),
\] 
and the pull back condition $\varphi^\ast(\Theta) =0$ gives rise to the first 
law of thermodynamics, i.e.
\begin{equation}
d\Phi =\sum_a I_a dE^a, \label{1st}
\end{equation}
together with the equations of states
\begin{equation}
I_a = \frac{\partial \Phi}{\partial E^a}. \label{2nd}
\end{equation}

It is known from standard texts on differential geometry that a contact 
manifold possesses a very special group of symmetries, i.e. the group of 
contact transformations. The group of Legendre transformations, i.e. 
\begin{align*}
\Phi = \tilde{\Phi} - \sum_{a\in \mathcal{I}'} \tilde{I}_a \tilde{E}^a, 
\quad 
E^a = - \tilde{I}_a, \quad 
I_a = \tilde{E}^a,
\end{align*}
for $a \in \mathcal{I}'$ with $\mathcal{I}=\{0,1,2,...,n-1\}$ and 
$\mathcal{I}' \subset \mathcal{I}$ is any subset therein, is a subset of 
the group of contact transformations, and H.Quevedo stressed very much 
on the invariance under this subgroup in his prescription of 
geometrothermodynamics \cite{Quevedo:2006p1979} 
\cite{Quevedo:2007p1262}. It is however not my intention to keep my 
eyes addicted to the group of Legendre transformations. Rather, the 
introduction of a Riemannian metric on $\mathcal{T}$ seems more 
attractive. The concrete form of the metric on $\mathcal{T}$ is not important. What is important is that the pull back of this metric naturally introduces a metric on the thermodynamic configuration space $\mathcal{E}$,  which, in its most general form, can be written as
\begin{equation}
ds^{2} = g_{ab}(E) dE^{a} dE^{b}. \label{dsmetric}
\end{equation}

Now one of the crucial part of this work turns up. There exists a large 
continuous group of hidden symmetries which keeps both
(\ref{1st}) and (\ref{dsmetric}) invariant. This hidden symmetry group
is the group ${G}$ of general coordinate transformations in the space $
\mathcal{E}$, i.e.
\begin{align*}
E^a &\rightarrow E^{\prime a} = E^{\prime a}(E^b), \quad \Phi 
\rightarrow \Phi'=\Phi,\\
dE^a &\rightarrow \left(\frac{\partial E^a}{\partial E^{\prime b}}\right) 
dE^{\prime b},\qquad
I_a \rightarrow \left(\frac{\partial E^{\prime b}}{\partial E^{a}}\right) 
I'_{ b}.
\end{align*}
In the above, the transformation law of $I_a$ is triggered by the relation 
(\ref{2nd}). 

Borrowing some terminologies from the general theory of relativity, $dE^a$ 
transform under the group ${G}$ as a contravariant vector, $I_a$ transform 
as a covariant vector, while $\Phi$ transforms as a scalar.

Notice that although in the above I am using geometric terminologies like 
manifold, Riemannian metric, contravariant and covariant vectors etc, 
no fundamental existence of space is assumed actually. As Verlinde has 
emphasized, the definition of thermodynamics does not need the existence 
of space. Space actually emerges as macroscopic consequences as some of 
the extensive variables (generalized displacements) in the space 
$\mathcal{E}$.  Therefore, the group $G$ is by now purely 
thermodynamic in nature, nothing 
relevant to spacetime symmetry has entered into play. 

The invariance under the group $G$ is totally a new 
observation which has not been discussed before. Considering the fact that 
there has already been quite some works \cite{F. Weinhold} 
\cite{G. Ruppeiner} assigning a Riemannian metric to 
the space of equilibrium states of thermodynamics, it is quite strange why 
this large group of isometries have not been discussed before in the 
context of thermodynamic geometries (for the case of Weinhold geometry, 
the isometry group is discussed in \cite{P. Salamon}, but the groups 
considered there is not the same as the one discussed here. Meanwhile, the 
deep implications implied by the symmetries is somehow not widely 
acknowledged). Anyway, the finding of this large group of 
hidden symmetries provides room for an 
interpretation of the principle of relativity as an emergent effect from 
thermodynamics, as will be seen below. 

First let me fix the choice of the thermodynamic potential $\Phi$ by 
identifying it as the total internal energy of the system. This breaks the Legendre 
symmetry but leaves the $G$ invariance unaffected. Meanwhile, this 
specific choice of thermodynamic potential implies that the entropy $S$ is 
among the extensive variables, let me give it the index $0$, i.e. $E^{0}=S$. 
Unlike the case of Weinhold and Ruppeiner, I did not associate the metric 
$g_{ab}(E)$ with the hessian of any specific thermodynamic function, so 
there is still enough freedom in choosing the signature of $g_{ab}(E)$. Let 
the signature be Lorentzian and let $E^{0}$ be the coordinate bearing the 
different signature from others. The $G$ symmetry allows to make different 
choices for the reference frames on $\mathcal{E}$. Near any point $X$ in 
$\mathcal{E}$, take a small neighborhood $U$ thereof, then using 
$G$-transformations one can always fix the metric (\ref{dsmetric}) on $U$ 
to a special form, i.e.
\begin{equation}
ds^{2} = \eta_{ab} dE^{a} dE^{b}, \label{dsinertial}
\end{equation}
where $\eta_{ab}$ is the Lorentzian metric in $n$ dimensions.
This is the analogue of taking a local inertial frame in the general theory of 
relativity, but now practiced purely in the framework of thermodynamics. 
Once the frame (\ref{dsinertial}) is taken, the $G$ symmetry is broken, but a residual subgroup $H$ survives, which consists of linear 
transformations among $E^a$, i.e.
\begin{equation}
E^a \rightarrow E^{\prime a} =\Lambda^a{}_b E^b + T^a,
\label{linear}
\end{equation}
in which $\Lambda^a{}_b$ belongs to the orthogonal group $SO(n-1,1)$ 
and $T^a$ are constants. Clearly eq.(\ref{1st}) is 
invariant under such a group of transformations, provided $I_a$ transform 
inversely under $H$. 

Readers may have already noticed the close analogue of eq.(\ref{linear}) 
with the Poincare group. To actually establish the relationship between the 
group $H$ and the Poincare group, extra input is needed, including the 
second law of thermodynamics and Verlinde's proposal for an emergent 
space. 

Consider first the second law of thermodynamics. There are many
presentations for the second law in the literature. For convenience, we take 
the following presentation:

Second law: {\em In any thermodynamic process connecting two 
equilibrium states of an isolated macroscopic system, the entropy does not 
decrease, \i.e.}
\[
dS \geq 0.
\]
This presentation can also be put in another way, i.e. {\em any equilibrium 
state of a given macroscopic system of lower entropy cannot be the 
consequence of a thermodynamic process of another equilibrium state of 
higher entropy.} Such a statement is reminiscent to the causality principle 
of relativistic physics and may be called the thermal causality principle.

Careful readers may have felt uneasy with the last paragraph. 
According to eq.(\ref{linear}), $S$ is not a scalar under the action of the 
group $H$. In other words, there is not a unique choice for the extensive 
variable $S$ in the presence of the symmetry group $H$. So the thermal 
causality principle must 
be formulated in an invariant way under the action of the group $H$. For 
this purpose we need an invariant quantity under the action of $H$, and 
the line element on the space $\mathcal{E}$ happens to fill this 
gap.

Geometrically the line element (\ref{dsinertial}) describes the invariant 
distance between two points in $U$. What is the thermodynamic meaning 
of such a distance? It is clear that the two points connected by $ds^{2}$ 
correspond to two distinct equilibrium states, so the line element between 
them must corresponds to a thermodynamic process evolving from one 
state to the other.  Now the following crucial question arises: given any two 
states $A, B$ in $U$, does the thermodynamic process connecting them 
always exist? To answer this question, let me first fix a reference frame on 
$U$ such that $S_{A} \leq S_{B}$. Then the second 
law implies that there can possibly be a thermodynamic process 
evolving from the state $A$ to the state $B$. However, such a process is 
not guaranteed to exist, since a change of frame on $U$ can 
spoil the inequality $S_{A} \leq S_{B}$. So, if $A$ and $B$ are such that 
$S_{A}< S_{B}$ in one frame but $S_{A}>S_{B}$ in some other frame, 
the presumed thermodynamic process evolving from $A$ to $B$ (or vise 
versa) should be excluded by the second law. In other words, such states 
$A$ and $B$ must not be causally connected via thermodynamic process. 
On the other hand, it is possible that for properly chosen states $A$ and $B
$, the inequality $S_{A} \leq S_{B}$ holds for all allowed choices of 
frames on $U$. In such cases, one cannot exclude the 
possibility that $B$ is the thermodynamic consequence of $A$ through 
some thermal process. It is not a hard practice to show that $S_{A} 
\leq S_{B}$ holds for all allowed choices of frames on $U$ if and only if 
$S_{A} \leq S_{B}$ holds in one frame on $U$ and the line element $ds_
{AB}^{2}$ between $A, B$ obeys $ds_{AB}^{2}\geq 0$. So, 
the thermal causality principle can be formulated as follows: 

Invariant presentation of the second law: {\em In any thermodynamic process connecting two equilibrium states of 
an isolated system, the line element $ds^2$ must be nonnegative.}

For this reason we can possibly call $ds$ the {\em proper entropy change} 
between the two states. Notice that I did not assign an arrow to the line 
element, so a thermodynamic process corresponds to both $ds_{AB} 
\geq0$ and $ds_{AB} \leq0$. 

Now let me follow Verlinde's idea and consider the whole universe as an 
emergent macroscopic system. This implies, among other things, that the 
spacial coordinates $X^i$ are among the 
extensive variables $E^a$ in the space $\mathcal{E}$. Moreover, since 
nothing is assumed to exist outside the 
universe, one can consider the universe as an isolated system,  i.e. the 
condition for the thermal causality principle hold, any physical 
thermodynamic process must obey $ds^{2} \geq 0$. 

Unlike Verlinde's original proposal for a single emergent dimension, I 
postulate here that {all the spacial dimensions are emergent}. With the 
emergent spacial coordinates as extensive variables, the first law (\ref{1st}) 
should be modified as follows:
\begin{equation}
dE = TdS - F_i dX^i  - pdV + \mu dN, \label{law}
\end{equation}
where the sum over $i$ extends through all spacial dimensions. Note that  
$F_i$ are emergent forces just like $p$ does, and $X^i$ and $V$ are all 
macroscopic quantities which need not have an microscopic origin. This is 
what the term emergence of space means. 

Since the spacial coordinates are identified as 
generalized displacements, the total number of such displacements must be 
an integer. This explains why 
the dimension of space is an integer. It is natural to assign a scaling  
dimension for each of the extensive variables. Doing so one sees that only 
when space is three dimensional and $S$ is proportional to the area of a 
holographic screen, the right hand side of eq.(\ref{law}) can be 
{\em democratic} (i.e. evenly distributed) between quantities of different 
scaling dimensions. In the above, $N$, the number of total microscopic 
degrees of freedom, is zero dimensional, $X^i$ 
are one dimensional, $S$ is two dimensional and $V$ is three dimensional.  
In this way the role of holographic principle in Verlinde's proposal and the 
implicit assumption of $3$ spacial dimensions are replaced by a single 
requirement of democracy between scaling dimensions. 

In a process with $dV=dN=0$, the action of the subgroup 
$P$ of $H$ leaving $V$ and $N$ invariant is isomorphic to the 
Poincare group in the subspace $\mathcal{M}$ of $\mathcal{E}$ spanned 
by $S$ and $X^i$. If I assume that $dS$ is proportional to the time 
elapsed between the two equilibrium states (and this has to be so, because 
in any reference frame, $S$ is a monotonic function of time $t$, thus $dS 
\propto dt$ locally, and the procedure of fixing the metric (\ref{dsinertial}) 
could fix the constant of proportionality to 1), then the relativistic causality 
principle naturally follow from the thermal causality 
principle.

What is the role of $H$ in the process involving non zero $dV$ or $dN$? 
The answer could also be very interesting. For instance, if a process 
involves non zero $dV$, then the group $H$ is bigger than $P$, i.e. the 
pull back line element $ds^2$ possesses a bigger isometry group. Such a 
group provides room for cosmological expansion, and the corresponding 
intensive variable $p$ could possibly play the role of dark energy. However, 
at present, such possibilities must be regarded as speculative, because much 
more work on the understanding of thermodynamic emergence of space  
has to be done before such speculations can be made more sounded. As 
for $dN \neq 0$, the process will involve production or destruction of 
particles and that is beyond the present discussion on the emergence of the 
principle of relativity.

\section{Discussions}

Using the hidden symmetry group of thermodynamics and the thermal 
causality principle it is found that the principle of relativity arises 
naturally as an emergent consequence of thermodynamics. Thermal 
causality is identified with the relativistic causality by requiring  
that the increase in entropy, $dS$, is proportional to the time elapse. This 
analysis resolves a number of problems left over from Verlinde's proposal 
for an emergent space and gravity. 

One may wonder why in the first place the principle of relativity could arise 
as an emergent consequence of thermodynamics which has long been 
known to be incompatible with special relativity regarded as a fundamental 
principle  theory. In particular, why is the total energy $E$ taken as the 
thermodynamic potential transform as a scalar under the emergent Poincare 
group, while in traditional special relativity, energy transforms as the zeroth 
component $p^0$  of the energy-momentum 4-vector. The answer is that  
$p^0$ and $E$ are completely different objects which should be not 
confused with each 
other. Actually, $p^0$ refers to the energy of certain microscopic degree of 
freedom, while $E$ counts the total energy of all microscopic degrees of 
freedom in the system, so by definition $E$ is an integral quantity in 
which the covariant and contravariant actions of Poincare group cancel 
completely. Due to the same reason, there is no reason to write $E=Mc^2$, 
because $Mc^2$ only represents the energy of a microscopic 
degree of freedom of mass $M$ at rest, while $E$ counts the energy of all  
microscopic degrees of freedom, and most of these are not at rest.

Another reason supporting the view point of not regarding $E$ as a 
component of relativistic $4$-vector comes from the new relationships 
between relativity and thermodynamics. If spacetime were fundamental and 
$E$ is the total energy of a thermodynamic system moving in space, then it 
seems inevitable that $E$ should change under relativistic change of 
coordinates. However, in the present picture, {\em space itself is emergent 
from thermodynamics}. The thermodynamic system no longer moves in 
space, and $E$ contains not only the energy of ordinary matter but also the 
energy of space. (The picture of space as emergent from 
thermodynamics implies that space itself can be heated, and thus stores 
energy. On this point, I am grateful to T.~Padmanabhan for bringing my 
attention to the paper \cite{Padmanabhan:2010} after the first versions of 
the present paper have appeared on arXiv.) We human beings have no 
previous 
knowledge at all on how the total energy of matter and space changes 
under coordinate changes in space and time.

In this article, only the linear subgroup of the hidden symmetries of 
thermodynamics is analyzed in detail. A more thorough analysis on the 
complete hidden symmetry group of thermodynamics should be considered  
later, which is expected to reveal how general relativity arises from 
thermodynamics. Whatever the details will be, I expect that a better 
derivation of Einstein equation should follow and Newton's law of gravity 
should arise as nonrelativistic limit of general relativity as usual, thus 
removing the puzzle of deriving nonrelativistic laws of gravity from relativistic 
formulas. 

The argument made in this article is very preliminary. Much more works are 
yet to be done. Among other things the hidden symmetry group $G$ of 
thermodynamics is bigger than the general coordinate transformations for 
general relativity provided in the first law of thermodynamics $dV$ and $dN$ 
are not simultaneously zero. The role of the extra symmetries is yet to be 
understood, and perhaps this will give novel insights into the general theory 
of relativity.

\section*{Acknowledgment} 

This work is supported by the National  Natural Science Foundation of 
China (NSFC) through grant No.10875059.


\providecommand{\href}[2]{#2}


\end{document}